\documentclass[12pt]{article}
\usepackage{geometry}                
\usepackage{graphicx}
\usepackage{amssymb}
\usepackage{epstopdf}
\title{Physics Pedagogy and Assessment in Secondary Schools in the U.S.}
\author{Gordon P. Ramsey, Melissa M. Nemeth and David Haberkorn \\
Loyola University Chicago, Chicago, IL 60626}
\date{}                                           

\begin{document}
\maketitle

\abstract{The objective of this project is to compare the effectiveness of teaching styles used in high 
school physics classes and the methods used to assess them. We would like to determine those 
approaches to physics at the high schools that work and those that do not work for students from 
different demographics. We sent out a survey to high school physics teachers in the U.S. midwest 
states, inquiring about student preparation, pedagogy in the classroom, assessment and professional 
development. We found that there are differences in the practices of physics teachers in all of these 
areas, depending on the school location, be it rural, suburban or urban. Our results enable us to report
on the most common successful practices in physics courses for these demographic areas.}

\section{Introduction}
The motivation behind this study is a desire to improve physics education, especially at the secondary school level. We believe that science is an essential component of education for all, and one of the key purposes of science education should be to promote scientific literacy and appreciation. 

Recent TIMSS \cite{TIMSS} and ÒNation at RiskÓ \cite{NAR} reports from the U.S. government indicate that our high school science students are behind most of the industrialized countries in the world. This has been attributed to such factors as teachers that are ill prepared to teach in science subjects (particularly physics), outdated curriculum and teaching methods that are ineffective. The purpose of this 
project is to isolate one of these factors, namely teaching methods (which may be coupled to curriculum) 
to determine those that have proven most effective for different populations of high school physics 
students. Similarly, we would like to know if any methods were not effective and possibly isolate the 
reasons for their ineffectiveness. There exist studies of various teaching methods in secondary physics, 
\cite{BW, Rennie, Linden} but none address the students for which these methods appear to be 
the optimum. That is where this study is unique and of interest to a wide audience of high school physics 
teachers. The assessment methods will also be studied to determine how teachers determine the 
effectiveness of their methods.

As physics provides a crucial link between mathematics and science, high school physics teachers are under constant pressure to deliver the best education possible. Our research aims to uncover current best practices in secondary physics education and make recommendations based on our key findings. With the knowledge that studentsÕ socioeconomic status and teachersÕ experience affects the way physics is taught, we surveyed teachers in the categories of demographics, student and teacher backgrounds, teaching practices, and assessment techniques. Using current education research, we created a measuring tool to rank and quantify responses in these categories. We used these numbers to quantify the key findings presented. Our main objective is to make recommendations of specific ways to make high school physics more engaging with the ultimate goal of ensuring higher student success in college and beyond. 

\subsection{U.S. Physics Education and Science Standards}
There are four major levels of education in the United States \cite{ISBE}: grammar school (K-5), middle school (6-8), high school (9-12) and college (13-16). Basic science education begins during the 
grammar school years. The divisions of biology, chemistry and physics are taught at the middle school
level. Although science curriculum varies in different states, some programs integrate the sciences
during the middle school years. The high schools have separate courses in biology, chemistry and 
physics, but ordering varies from state to state. Each state has their own science standards, but there
are movements to introduce and test nationwide standards. Studies like this one may lend insight on
how current practices can help meet and exceed the standards, regardless of the level at which they are
imposed. 

There are three major geographic distributions of school districts in the U.S.: rural, suburban and urban.
Rural schools are prevalent in the largest geographical area of the U.S. They lie outside the larger cities and suburbs. On average, there are one to two high school physics teachers per district. Most schools
are public with few private schools. The school sizes vary, but average about 700 students per school. 
Suburban schools are those that lie outside the limits of most larger cities and may encompass the
counties surrounding the city itself. They consist of a mix of public and private schools. The student
population averages about 1650 students per school with three to seven physics teachers per school. 
Urban schools are those within the city limits of cities with population of 100,000 or more. They also consist of a mix of public and private schools. Typically there are many schools within the geographic 
area, so there are fewer students per school at about 1400. Depending upon the city district, there are 
on average one to four physics teachers per school.  

Typical science standards for secondary schools in the midwest include the following elements: 
\begin{enumerate}
\item Applications for learning (inquiry)
\item Formulate and solve problems (concepts)
\item Interpret information and ideas (principles)
\item Use appropriate instruments (design)
\item Connect ideas among learning areas (STS)
\item Common Core standards in science
\end{enumerate}
Accordingly, our survey asked questions on guided inquiry, pedagogy, technology, engaging curriculum, homework and group work, communication and assessment. We purposely did not ask
specifically about teaching to standards, since the focus of our work is what actually takes place in and
out of the classroom. This paper is a report of work in progress. It outlines the key results of our survey
and summarizes the classroom practices that are typical in high school physics courses for rural,
urban and suburban areas in the Midwest portion of the U.S.

\section{Overview of research methods}
\subsection{Survey overview and key definitions}
\subsubsection{Overview of the Survey}

Survey data was obtained from high school physics teachers in seven Midwest states. We asked what methods they have found to be effective. The effectiveness of these is substantiated by appropriate assessment (e.g., grades or standardized test scores). The data were compared with the demographics 
of their students, such as whether the school is located in a rural, suburban or urban area, as this is an 
important factor in their approach to teaching physics and their degree of success. 

The survey consists of fifty questions, divided into six parts:
\begin{enumerate}
\item Demographics
\item Student Preparation
\item Pedagogy
\item Communications and other skills
\item Assessment
\item Professional Development.
\end{enumerate}
The demographic information is used to determine backgrounds, experience that our respondents have and the location and type of school where they teach. These variables may help to distinguish if 
background and teaching experience are variables that affect the classroom and assessment practices
in high school physics classes. Much of our analysis tries to distinguish practices for different bodies of 
students. Therefore, we divided the schools into five groups, separated by location (rural, suburban and 
urban) and type (public or private), since each cohort experiences different constraints and standard 
practices. The aim is to be able to make recommendations for each group, based upon their particular 
cohort of students.

It is well known from science education studies that students enter physics courses with conceptions
that are not accepted by science, or ``misconceptions". These are typically firmly embedded in their
understanding of science and must be overcome for the students to have a fundamental understanding
of physical concepts. \cite{Hammer} The Student Preparation section of the survey attempts to determine
the misconceptions that students of each cohort have when the enter a physics class. We follow up by
asking what the teachers are doing to address these misconceptions. 

The section on Pedagogy reflects one of the main focuses of the study. We ask about techniques used
in the classroom, curriculum, technology and making the experience relevant to the students. This is
the section that is most pertinent to the science standards listed in the last section. We
correlated this with background, experience and other teaching practices for each respondent. The
results of this analysis can be valuable for teachers to improve their effectiveness for each cohort of 
students. By comparing what others have been successful in implementing, it is hoped that teachers
in a similar situation can improve their students' experience in physics classes. 

It is important to frequently communicate with students, particularly on their progress toward learning
goals. \cite{Rennie} Communications can serve two purposes: (i) to inform the student on how they can 
improve their performance and (ii) to inform the teacher about class weaknesses so that they can be 
properly addressed. Thus, we inquired about the methods and frequency of communication with 
students and how it affects teaching practices. 

The Assessment section is the other main emphasis of our study. This determines how the teachers
measure the effectiveness of the pedagogical approach that they outlined in the previous section of
the survey. We asked about various methods used for assessment and how the information is used
for determining effectiveness and adjusting classroom practices to achieve the course goals. 

Finally, morale and school environment of a teacher are elements that can play a vital role in their
effectiveness and motivation to continue teaching. We wanted to determine the extent of mentoring
and collaboration for the teachers of each cohort to suggest ways that teachers can be more effective
in a given learning environment. Both the presence of collaborators and involvement in professional
organizations can contribute effectively to this motivation. We concluded the survey with related 
questions.

\subsubsection{Introduction to Quantitative Analysis}
In an effort to quantify the different aspects of classroom engagement level, we asked several questions about teachersÕ curricula and classroom management. As detailed below, created a set of quantitative analysis numbers to describe engagement levels in different ways: the Nominal Response Number, the Engagement Number and the Curriculum Relevancy Number. Since assessment is an important aspect 
of determining the effectiveness of the pedagogy, we assigned an Assessment Number to each respondent. These numbers were based on research that shows the more different ways teachers engage and assess their students, the more the students will learn.
\begin{description}
\item[Numerical Response Number (NR$\#$)] The NR$\#$ is a weighted average of the percentage use of these 5 methods in the classroom: lecture, demonstration, discussion, problem solving, and laboratory, in order of student involvement with lecture being the least. The lecture is assigned a value of one and the laboratory five, with the others represented by the integers in between. To find the NR$\#$, the decimal equivalent of the percentages of each method are multiplied by the integer and summed. Thus, $$NR\# = \sum_{methods} \frac{\%}{100}\times \mathrm{integer\> value\> of\> method}.$$ The NR$\#$ has a range of from 1 to 5 and is a rough measure of the degree of student involvement in the classroom. 
\item[Engagement Number (E$\#$)] The E$\#$ is a sum of the number of engaging methods used  from this list: labs, video, demonstration, group work, feedback systems, projects, online communities. The 
E$\#$ is an indication of the variety of engagement methods used in the classroom, without regard to weighting the effectiveness of each.
\item[Assessment Number (A$\#$)] The A$\#$ is a sum of the number of formal assessment tools used from this list:
\begin{enumerate}
\item $\it{Traditional}$: tests, quizzes
\item $\it{Portfolio}$: journals, reflections, lab notebooks, projects, presentations, demonstrations, creative products.
\end{enumerate}
The A$\#$ indicates the variety of assessments that the teacher uses as part of their courses. 
\item[Curriculum Relevancy Number (CR$\#$)] The CR$\#$ is a sum of the number of methods used to make the curriculum relevant to studentsÕ lives from the following: articles, practical experiments, live demonstrations, realistic physics problems, guest speakers, field trips, social networking). This is a measure of the teacherÕs efforts to bring practical applications to their curriculum. 
\end{description}
Many of our findings were correlated with these numerical quantities. As there are many correlations
in our analysis, we include only a subset of these to be reported here.  

\subsection{Demographics}
The demographical information of teachers with valid responses is shown in 
Table \ref{demographicsgender}, where we separate the categories by gender. This is to ensure that we
have adequate diversity in the responses. The secondary school physics teaching experience 
categories are defined as ``New" ($1-3$ years), ``Intermediate" ($4-6$ years) and ``Experienced"
(more than six years). A majority of the respondents are experienced teachers, which gives us an
indication of what teachers are doing that have possibly tried different approaches in and outside of
the classroom. The relative numbers of public and private schools represented are consistent with
the actual percentages for each of the three location categories. Table \ref{demographicsexperience}
shows the relative experience and physics background training for the three location categories.
These are all consistent with each other, showing that we have targeted experienced teachers with
background training equivalent to a major or very strong minor in physics.

We sent just over 1300 surveys by e-mail to high school teachers in seven Midwest states. After $2-3$
months, we received $93$ responses that completed the demographic data (a rate of $7.2\%$). 
A total of $79$ completed the entire survey. All of these teachers were present or former members of the
American Association of Physics Teachers (AAPT). We chose this cohort, since these would be
teachers that are likely to employ a diversity of teaching techniques, involve the students in their own
learning and be slightly more experienced in the high school physics classroom. We realize that this
is not all inclusive, but future work will include a much broader base of teachers. 
Table \ref{demographicsgender} gives a demographic distribution by gender, including new, 
intermediate and experienced teachers. Most of the teachers are experienced, as we expected from this 
cohort. A large majority is in public school systems, since these are a major portion of rural and urban 
schools. The ``average" backgrounds of the respondents are shown in 
Table {demographicsexperience}. The ``semester hours of physics" category inquired about how much 
physics training the teacher had in college. Between ten and twenty would be considered a minor and 
more than twenty, a major in physics. The college training and teaching experience are consistent for all 
categories of schools, indicating a more homogeneous group. 
\begin{table}[htdp]
\caption{Gender Distribution of Demographics in total numbers}
\begin{center}
\begin{tabular}{|c|c|c|c|c|c|c|}
\hline
\hline
Gender & New & Intermed. & Experienced & Public & Private & Sem Hr Phys \\
\hline
\hline
Female & 1 & 2 & 22 & 18 & 5 & 20 \\
\hline
Male & 3 & 10 & 58 & 53 & 11 & 30 \\
\hline
\end{tabular}
\end{center}
\label{demographicsgender}
\end{table}%

\begin{table}[htdp]
\caption{Backgrounds of Respondents}
\begin{center}
\begin{tabular}{|c|c|c|c|c|}
\hline
\hline
Category & Rural & Suburban & Urban & Overall \\
\hline
\hline
Total number & 24 & 48 & 22 & 92 \\
\hline
Average years teaching & 19 & 17 & 20 & 18 \\
\hline
Average sem. hrs. phys. & 30 & 26 & 25 & 27 \\
\hline
\end{tabular}
\end{center}
\label{demographicsexperience}
\end{table}%

\section{Key findings}
\subsection{Pedagogical Data}
SUMMARIZE the sections of pedagogy and significance of the information. pedagogy in the classroom, student misconceptions and ways to 
overcome, making curriculum engaging and relevant, use of technology, ways to improve courses, 

\subsubsection{Preparation of entering Students}
In science education research, it is accepted that students come to courses with conceptions that differ from scientists' and must be addressed in instruction. These  "misconceptions" (i) are strongly held; (ii) differ from expert conceptions; (iii) affect how students understand natural phenomena and scientific explanations; and (iv) must be overcome for students to achieve expert understanding. \cite{Hammer} 
One of the initial concerns of a physics teacher is the background with which the students enter the
course. We asked about the key misconceptions with which students typically enter the course and
what the teachers do to bring their students to a satisfactory level for the course material. 
Table \ref{misconceptions} indicates the areas in which misconceptions exist, separated by the three
locations (rural, suburban and urban) and by public versus private schools. Lack of understanding
physics concepts is the leading area for all categories. This may come from not having a prior course
in physics. Physical science courses in middle schools often emphasize the biological sciences at
the expense of the physical sciences. Weak math backgrounds are typical of about two-thirds of the
students. This tends to be higher in private schools, although it is not clear why. The results in the other 
categories of units, nature of science and matter are dependent upon the location. Students in rural and
urban schools typically do not have as strong backgrounds as more affluent suburban schools.  
The choices for methods to overcome this lack of preparation included science review, math review, 
embed misconceived topics into the course work, individual assistance, separate tutoring, adjusting the curriculum and adjusting the pace of the course).
\begin{table}[htdp]
\caption{Types of Misconceptions/Weaknesses (percent of respondents)}
\begin{center}
\begin{tabular}{|c|c|c|c|c|c|}
\hline
\hline
Demographic & Units & Nature of Science & Math & Matter & Physics concepts \\
\hline
\hline
Rural & 39 & 65 & 65 & 57 & 96 \\
\hline
Suburban & 50 & 41 & 70 & 43 & 93 \\
\hline
Urban & 70 & 60 & 70 & 43 & 93 \\
\hline
Public & 46 & 51 & 68 & 51 & 94 \\
\hline
Private & 75 & 56 & 81 & 38 & 88 \\
\hline
\end{tabular}
\end{center}
\label{misconceptions}
\end{table}%

\begin{table}[htdp]
\caption{Methods to overcome the lack of preparation (percent of respondents)}
\begin{center}
\begin{tabular}{|c|c|c|c|c|c|c|c|}
\hline
\hline
Demographic & Sci rev & Math rev & Embed & Assist & Tutor & Adjust curr. & Pace \\
\hline
\hline
Rural & 30 & 39 & 87 & 70 & 22 & 57 & 43 \\
\hline
Suburban & 25 & 20 & 84 & 73 & 27 & 50 & 34 \\
\hline
Urban & 15 & 35 & 80 & 65 & 15 & 50 & 15 \\
\hline
Public & 24 & 31 & 82 & 69 & 23 & 54 & 38 \\
\hline
Private & 25 & 19 & 94 & 75 & 25 & 44 & 6 \\
\hline
\end{tabular}
\end{center}
\label{adjustment}
\end{table}%
The most common practices for those teachers that have the most interactive classrooms (indicated by 
the NR$\#$) and most engaging classes (indicated by the E$\#$) are embedded review, individual 
instruction (assistance) and adjusting the curriculum. These are shown in 
Table \ref{topmisconceptionpractices}. 
\begin{table}[htdp]
\caption{Addressing misconceptions-top teachers' practices}
\begin{center}
\begin{tabular}{|c|c|c|c|}
\hline
\hline
Top respondents & Embedded Review & Individual instruction & Adjust curriculum \\
\hline
\hline
NR $\ge 2$ (avg=$1.94$) & 79 & 74 & 54 \\
\hline
E $\ge 4$ (avg=$3.98$) & 86 & 73 & 57 \\
\hline
\end{tabular}
\end{center}
\label{topmisconceptionpractices}
\end{table}%

\newpage
\subsubsection{Engaging and Relevant Curriculum}
Studies have shown that students are more motivated and learn more when the curriculum is made
engaging and relevant. \cite{MPT} There are various ways to accomplish the goal of making the
classroom engaging, including but not limited to:
\begin{itemize}
\item computer simulations
\item laboratory experience
\item demonstrations 
\item group discussion and problem solving
\item frequent feedback on work
\item project work, and
\item online resources.
\end{itemize}
We asked teachers about the ways that they make their classes more engaging. Table \ref{engagement}
indicates the percentages of respondents that use these techniques in their classes. Laboratories and
demonstrations are by far the most frequently used in all locations. This is followed by group work and
computer simulations. Rural schools have access to computers, but not as much demonstration
equipment available. This may explain the difference in those columns for these schools. Since we did not ask about the amount of labs that were performed, it is not clear from the survey results how diverse the lab equipment is typically available to the rural schools. The average engagement number for the
cohort was about $4.4$, indicating the number of ways that teachers make the classroom engaging.
\begin{table}[htdp]
\caption{Ways to make curriculum engaging (percent of respondents)}
\begin{center}
\begin{tabular}{|c|c|c|c|c|c|c|c|}
\hline
\hline
Demographic & Simulations & Labs & Demos & Groups & Feedback & Projects & Online \\
\hline
\hline
Rural & 63 & 92 & 63 & 67 & 17 & 42 & 8 \\
\hline
Suburban & 39 & 85 & 67 & 72 & 28 & 24 & 7 \\
\hline
Urban & 41 & 82 & 77 & 68 & 14 & 41 & 14 \\
\hline
Public & 47 & 87 & 68 & 69 & 21 & 32 & 9 \\
\hline
Private & 44 & 81 & 75 & 75 & 31 & 38 & 6 \\
\hline
\end{tabular}
\end{center}
\label{engagement}
\end{table}%

A Scientific American feature article by Gibbs and Fox \cite{GF} states Òthe false crisis in science education masks the sad truth that the vast majority of students are taught science that is utterly irrelevant to their livesÓ. In addition to making the classroom engaging, it is important to attach relevance to the physics topics. Teachers can make the material relevant to their students' lives in a number of ways, including:
\begin{itemize}
\item articles relating to the physics topic
\item association to life experiences
\item demonstrations with simulations and modeling
\item experiential examples with realistic numerical values
\item guest speakers
\item field trips, and
\item online resources.
\end{itemize}
Table \ref{relevance} indicates the percentage of respondents that use each of these methods to make
the material relevant. The most prevalent of methods include life experiences and related numerical examples. Demonstrations are used extensively, except in rural areas where less equipment is typically available. Articles and online resources are used by about one half of the cohort. Speakers and trips are 
less used due to the time and monetary costs of these methods. It is apparent that more connections of
physics topics to the ``real world" are more effective in making physics relevant to students' experiences.
This is not surprising, but confirms our expectations with actual data.

\begin{table}[htdp]
\caption{Making curriculum relevant (percent of respondents)}
\begin{center}
\begin{tabular}{|c|c|c|c|c|c|c|c|}
\hline
\hline
Demographic & Articles & Life Exp & Demos & Examples & Speakers & Trips & Online \\
\hline
\hline
Rural & 46 & 75 & 46 & 71 & 21 & 25 & 63 \\
\hline
Suburban & 43 & 74 & 68 & 72 & 13 & 38 & 49 \\
\hline
Urban & 55 & 86 & 77 & 64 & 9 & 32 & 45 \\
\hline
Public & 45 & 77 & 62 & 71 & 13 & 35 & 49 \\
\hline
Private & 50 & 81 & 75 & 63 & 19 & 25 & 63 \\
\hline
\end{tabular}
\end{center}
\label{relevance}
\end{table}%

\subsubsection{Role of Technology in the Classroom}
In the past three decades, the role of technology in the classroom has greatly evolved. In many classrooms, technology plays a primary role in learning physics. We asked the teachers about their use of PowerPoint presentation, simulations, Web resources, clickers, Web projects, video analysis and lab 
interfacing equipment. Table \ref{technology} indicates the percent of respondents that use these tools
in their classes. We found that teachers in our survey were more likely to use technology in the 
classroom if one or more of the following is true:
\begin{itemize}
\item the more semester hours of physics they had taken in college
\item the higher their Engagement Number (E$\#$)
\item the more ways they check for their effectiveness (A$\#$)
\item the higher their Curriculum Relevancy Number (CR$\#$).
\end{itemize}
Rural suburban and urban schools all showed similar use of technology in the classroom, showing that types of schools generally have equal access to technological resources. 

\begin{table}[htdp]
\caption{Use of technology (percent of respondents)}
\begin{center}
\begin{tabular}{|c|c|c|c|c|c|c|c|}
\hline
\hline
Demographic & PowerPoint & Sims & Web & Clickers & Web project & Video & Lab eqmt. \\
\hline
\hline
Rural & 42 & 71 & 33 & 8 & 4 & 46 & 75 \\
\hline
Suburban & 45 & 62 & 36 & 21 & 6 & 49 & 77 \\
\hline
Urban & 36 & 73 & 36 & 5 & 9 & 45 & 68 \\
\hline
Public & 44 & 65 & 32 & 13 & 6 & 45 & 75 \\
\hline
Private & 31 & 75 & 50 & 19 & 6 & 56 & 69 \\
\hline
\end{tabular}
\end{center}
\label{technology}
\end{table}%
The top four uses: are lab interface equipment, computer simulations, video analysis of phenomena and 
PowerPoint presentations. There is a significant increase in technology use for experienced versus new 
or intermediate teachers. 

\subsubsection{Improving Courses}
In light of all the pedagogical information, we wanted to determine what factors the teachers considered
important in improving their courses. The gives an indication of those elements that are not as strong
as the teachers would like for their courses. The choices include: smaller classes, more lab equipment,
larger budget, course development time and more technology. The results are shown in 
table \ref{improve}. The most desired element across the board was more time to further develop their
courses. Suburban schools tend to have large class sizes due to the numbers of students in the school 
that take physics (usually a larger percentage in the suburbs), so smaller class size was important in
those schools. Lab equipment is more in need at the rural schools due to smaller budgets per school.
\begin{table}[htdp]
\caption{Improving courses (percent of respondents)}
\begin{center}
\begin{tabular}{|c|c|c|c|c|c|}
\hline
\hline
Demographic & Smaller class & Lab eqmt & Budget & Devt time & Technology \\
\hline
\hline
Rural & 13 & 46 & 38 & 79 & 38 \\
\hline
Suburban & 53 & 23 & 28 & 51 & 17 \\
\hline
Urban & 36 & 18 & 23 & 55 & 9 \\
\hline
Public & 40 & 29 & 30 & 62 & 25 \\
\hline
Private & 31 & 25 & 25 & 44 & 0 \\
\hline
\end{tabular}
\end{center}
\label{improve}
\end{table}%

\subsection{Assessment}
Assessment serves the purpose of learning and is consistent with and complementary to good teaching. Teachers use a variety of assessment procedures to recognize where students are located in their development and plan learning experiences that move students toward desired learning 
outcomes. \cite{Rennie} We surveyed teachers on multiple aspects of their assessment practices to uncover the most common approaches to student evaluation and its purpose in the average high school physics teacherÕs classroom. In our study, teachers who gave more types of assessments received higher Assessment numbers. We then compared the teachers with the highest numbers to the average survey data to make recommendations.

Various assessment tools can be used to recognize students' progress. Tests fall into various categories,
including unit level, cumulative tests incorporating many units and standardized tests, which tend to be
more comprehensive. Grades are also based upon many units and cover a larger time period than unit
tests.Homework gives periodic feedback on how students are understanding the material. Projects
tend to occur at the end of a series of units or the end of a course, where many concepts are combined
to carry out the project. Private conversations with students tend to be a shorter range feedback
mechanism and are designed to help individual students, especially those that have more trouble
with understanding. Corresponding with the assessment tools, the most popular feedback methods
to follow up the assessment are written notes (for labs and homework), verbal discussion and forms of technology such as conversations or remedial tutorials. Table \ref{assess} shows the percent of
respondents that use these tools for assessment. 

\begin{table}[htdp]
\caption{Assessing effectiveness (percent of respondents)}
\begin{center}
\begin{tabular}{|c|c|c|c|c|c|c|c|}
\hline
\hline
Demographic & Grades & Std tests & Unit tests & Projects & Cum tests & HW & Converse \\
\hline
\hline
Rural & 46 & 8 & 92 & 25 & 38 & 54 & 71 \\
\hline
Suburban & 51 & 40 & 85 & 26 & 49 & 68 & 72 \\
\hline
Urban & 55 & 41 & 82 & 36 & 41 & 59 & 68 \\
\hline
Public & 48 & 29 & 87 & 27 & 42 & 60 & 74 \\
\hline
Private & 63 & 50 & 81 & 31 & 56 & 75 & 56 \\
\hline
\end{tabular}
\end{center}
\label{assess}
\end{table}%
In the verbal part of the survey, we asked how feedback from assessment is used to adjust elements
of the course. Most teachers said they gave assessments to assess what the students 
know about the content. Twenty-five percent said they used their teaching/learning environment to help 
students take responsibility for their own learning. When asked what ways teachers use their 
assessments to adjust, most teachers in urban areas responded that they adjust their curriculum. Most 
rural teachers reported that they then re-teach the topic with the assessment results in mind (this was the 
second most popular option for urban and suburban teachers). Most suburban areas adjusted their 
teaching style. No urban teachers reported that they used assessments to adjust for the future. 

\section{Discussion and Recommendations}
Our survey reflects the self reporting of how high school physics teachers structure their classes to
achieve the best results. The average data indicate what most teachers feel are best practices in their
courses. It is clear that these more experienced teachers reflect on their teaching practices and the
effectiveness of their courses. This validates the structure and content of the survey and our approach
to extract appropriate information about common teaching practices and effectiveness. 

Although students in the three demographic regions enter courses with different misconceptions, the
most popular methods used to overcome these weaknesses are fairly common across the board. This
indicates that these general approaches seem to work. See Tables 
\ref{misconceptions}-\ref{topmisconceptionpractices}. It is important to make the curriculum both
engaging and relevant to students' experiences. There are differences in the emphases on these
techniques, depending upon the location, but the most common (labs demonstrations and group work
for engagement and examples for relevance) are comparable for all areas. Refer to Tables
\ref{engagement}-\ref{relevance}. The use of technology is similar for all teachers, since most have 
access to some degree of equipment (Table \ref{technology}). However, there are significant differences 
in rural versus suburban an urban desires for improvement of the courses (Table \ref{improve}).
Unit tests, homework and conversations were the most popular techniques to assess student progress
(Table \ref{assess}).

It is clear from these results that schools in rural, suburban and urban areas have somewhat different
approaches to pedagogy, depending upon the background of their students and the resources that
they have available. This leads to teachers in these areas having different needs to improve their
approaches to the subject. 

We have presented a brief summary of a larger body of work in progress to determine common practices
in rural, suburban and urban high school physics courses. Our future plans consist of expanding the 
survey to a larger sample of teachers, outside the Midwest to encompass the geographical areas of the
northeast, east, south and west parts of the U.S. This will also include non-AAPT members in public and 
private schools). This will allow us to expand the analysis for a larger sample and perform a study of
differences in practices for these demographic areas. The recommendations we make will be based 
upon a wider spectrum of results, which could be valuable on a wide scale and of use to many more
high school physics teachers. \\

{\bf Acknowledgement} \\
We thank those who anonymously took the time to fill out our survey and those who gave us feedback to improve the format and content.


\begin{thebibliography}{100}
\bibitem{TIMSS} TIMSS, (2007). Third International Mathematics and Science Study report, 
http://nces.ed.gov/TIMSS/
\bibitem{NAR} Nation at Risk (1983). Nation at Risk report, http://www.ed.gov/pubs/NatAtRisk/index.html 
\bibitem{BW} Black, P. and Wiliam, D. (1998), Assessment and Classroom Learning. Education, 5, (1), 7-98, p. 7
\bibitem{Rennie} Rennie, LŽonie J.; Goodrum,  Denis and Hackling, Mark (2001). Science Teaching and Learning in Australian Schools: Results of a National Study. Res. In Sci. Educ., 31 455-498, 2001
\bibitem{Linden} Lindenfeld, Peter, (1985). A survey of high-school physics teachers in New Jersey. AmJ.Phys., 53, 1065-1069 
\bibitem{ISBE} ISBE (2012) Illinois State Board of Education, http://www.isbe.state.il.us 
\bibitem{Hammer} Hammer, David (1996). More than misconceptions: Multiple perspectives on student knowledge and reasoning, and an appropriate role for education research. Am.J.Phys., 64, 1316-1325 
\bibitem{MPT} McWilliam, Erica; Poronnik, Philip and Taylor, Peter (2008).  Re-designing Science Pedagogy: Reversing the Flight from Science. J. Sci. Educ. Technol. 17: 226 
\bibitem{GF} Gibbs, W. W. and Fox, D. (1999). The False Crisis in Science Education. Sci. Am., Oct., pp. 87-92 
\end{thebibliography}
\end{document}